\documentstyle[12pt]{article}

\newcommand{\eqn}[1]{(\ref{#1})}

\newcommand{\gc}{g^{C}}
\newcommand{\gd}{g^{D}}
\def\(#1{ ^{(#1)} }

\def\e{{\rm e}}

\def\beq{\begin{equation}}
\def\eeq{\end{equation}}
\def\be{\begin{equation}}
\def\ee{\end{equation}}
\def\bea{\begin{eqnarray}}
\def\eea{\end{eqnarray}}
\def\bd{\begin{displaymath}}
\def\ed{\end{displaymath}}

\makeatletter
\newdimen\normalarrayskip              % skip between lines
\newdimen\minarrayskip                 % minimal skip between lines
\normalarrayskip\baselineskip
\minarrayskip\jot
\newif\ifold             \oldtrue            \def\new{\oldfalse}
\def\arraymode{\ifold\relax\else\displaystyle\fi} % mode of array entries
\def\@arrayskip{\ifold\baselineskip\z@\lineskip\z@
     \else
     \baselineskip\minarrayskip\lineskip2\minarrayskip\fi}
\def\@arrayclassz{\ifcase \@lastchclass \@acolampacol \or
\@ampacol \or \or \or \@addamp \or
   \@acolampacol \or \@firstampfalse \@acol \fi
\edef\@preamble{\@preamble
  \ifcase \@chnum
     \hfil$\relax\arraymode\@sharp$\hfil
     \or $\relax\arraymode\@sharp$\hfil
     \or \hfil$\relax\arraymode\@sharp$\fi}}
\def\@array[#1]#2{\setbox\@arstrutbox=\hbox{\vrule
     height\arraystretch \ht\strutbox
     depth\arraystretch \dp\strutbox
     width\z@}\@mkpream{#2}\edef\@preamble{\halign \noexpand\@halignto
\bgroup \tabskip\z@ \@arstrut \@preamble \tabskip\z@ \cr}%
\let\@startpbox\@@startpbox \let\@endpbox\@@endpbox
  \if #1t\vtop \else \if#1b\vbox \else \vcenter \fi\fi
  \bgroup \let\par\relax
  \let\@sharp##\let\protect\relax
  \@arrayskip\@preamble}
\makeatother

\newlength{\extraspace}
\newlength{\extraspaces}
\begin{document}

\renewcommand{\footnotesize}{\small}

\addtolength{\baselineskip}{.8mm}

\thispagestyle{empty}

\begin{flushright}
\baselineskip=12pt
{\sc OUTP}-98-20P\\
hep-th/9803092\\
\hfill{  }\\ March 1998
\end{flushright}
\vspace{.5cm}

\begin{center}

\baselineskip=24pt

{\Large\bf{Gradient Flow in \\Logarithmic Conformal Field Theory}}\\[15mm]

\baselineskip=12pt

{\sc Nick E. Mavromatos\footnote{PPARC Advanced Fellow (U.K.).\\ E-mail: {\tt
n.mavromatos1@physics.oxford.ac.uk}}} {\sc
and Richard J.\ Szabo\footnote{Work supported in part by PPARC (U.K.).\\
E-mail: {\tt r.szabo1@physics.oxford.ac.uk}}}
\\[5mm]
{\it Department of Physics -- Theoretical Physics\\ University of Oxford\\ 1
Keble Road, Oxford OX1 3NP, U.K.} \\[15mm]

\vskip 1.5 in

{\sc Abstract}

\begin{center}
\begin{minipage}{15cm}

We establish conditions under which the worldsheet $\beta$-functions of
logarithmic conformal field theories can be derived as the gradient of some
scalar function on the moduli space of running coupling constants. We derive a
renormalization group invariant version of this function and relate it to the
usual Zamolodchikov $C$-function expressed in terms of correlation functions of
the worldsheet energy-momentum tensor. The results are applied to the example
of D-brane recoil in string theory.

\end{minipage}
\end{center}

\end{center}

\noindent

\vfill
\newpage
\pagestyle{plain}
\setcounter{page}{1}
\stepcounter{subsection}

Logarithmic conformal field theories~\cite{gurarie} are extensions of
conventional conformal field theories which have in recent years emerged in a
number of interesting physical problems in condensed matter physics~\cite{cm},
string theory~\cite{recoil}--\cite{lizzi}, and nonlinear
dynamical systems~\cite{nl}. They are characterized by the fact that
their Virasoro operator $L_0$ is not diagonalizable, but rather admits a
Jordan cell structure whose basis yields the set of logarithmic operators of
the
theory. The non-trivial mixing between these operators leads to
logarithmic divergences in their correlation functions. Such theories therefore
exhibit logarithmic scaling violations on the worldsheet. Nonetheless, it is
possible to deal rigorously with the Jordan cell structures of these theories
and classify their conformal blocks to some extent as in ordinary conformal
field theories.

Theories involving logarithmic operators lie on the border between conformal
field theories and generic two-dimensional quantum field theories. They
can be viewed as deformations of conformally-invariant theories whose critical
exponents are in general not scale invariant. Being rather subtle perturbations
away from the fixed points in the space of two-dimensional field theories, it
is natural to ask to what extent the conventional formalisms for studying the
properties of conformal field theories can be extended when logarithmic
operators emerge. In this letter we shall discuss the r\^ole played by
logarithmic conformal field theories in the geometry of the moduli space of
two-dimensional renormalizable field theories. In particular, we examine in
these cases an analog of the Zamolodchikov $C$-function~\cite{zam} which
interpolates among two-dimensional quantum field theories along the
trajectories of the renormalization group and whose stationary points coincide
with the fixed points of these flows. At these points the $C$-function is the
central charge of the resulting conformal field theory.

For a generalized $\sigma$-model on a Riemann surface $\Sigma$ with worldsheet
energy-momentum tensor $T$, the Zamolodchikov $C$-function \cite{zam}
\beq\new{\begin{array}{c}
\Phi[g]=\int_\Sigma d^2z~\left(2z^4\left\langle T_{zz}(z,\bar
z)T_{zz}(0,0)\right\rangle_g-4z^3\bar z\left\langle T_{zz}(z,\bar z)T_{z\bar
z}(0,0)\right\rangle_g\right.\\\left.-\,6z^2\bar z^2\left\langle T_{z\bar
z}(z,\bar z)T_{z\bar z}(0,0)\right\rangle_g\right)\end{array}}
\label{Cfundef}\eeq
can be regarded as an effective action in the space of two-dimensional
renormalizable field theories on $\Sigma$~\cite{tseytl,mm}. The expectation
values in
\eqn{Cfundef} are taken in a conformal field theory deformed by vertex
operators with associated coupling constants $g^I$. The corresponding classical
equations of motion are the gradient flow property of the $C$-function
which for unitary theories leads to the $C$-theorem
\beq
\frac{d\Phi}{dt}=-12\,\sum_{I,J}\beta^IG_{IJ}\beta^J\leq0
\label{Ctheorem}\eeq
where $t$ is the renormalization group parameter, $\beta^I[g(t)]=dg^I(t)/dt$
are the worldsheet $\beta$-functions, and $G_{IJ}$ is the positive-definite
Zamolodchikov metric on the moduli space of running coupling constants
$g^I(t)$. The $C$-function therefore decreases monotonically under
renormalization group transformations, thereby providing an analog of an
entropy function, and the $C$-theorem suggests a geometrical picture for the
equivalence of conformal invariance conditions and background string field
equations of motion.

In the following we shall examine conditions under which there exists a
suitable variant of the $C$-function whose gradient flows coincide with the
worldsheet $\beta$-functions associated with the perturbation of
conformally-invariant quantum field theories by logarithmic operators. We will
show that with a proper covariantization of the $\beta$-function equations
using the Zamolodchikov metric, there are indeed conditions under which such a
scalar
function exists. This is in contrast to the approach of \cite{r-ts} which used
non-covariantized $\beta$-functions, for which there is no gradient flow for
the coupling constants in logarithmic conformal field theories. The method we
shall employ is an extension of the approach of \cite{mms}, where by generic
scaling arguments it was argued that in conventional $d$-dimensional scale
invariant quantum field theories a gradient flow for the renormalization group
equations is always possible, within certain classes of renormalization
schemes.

We consider the simplest case of a two-dimensional logarithmic conformal field
theory that has a $2\times2$ Jordan block in which the logarithmic operators
$C$ and $D$ have equal conformal dimension $(\Delta,\overline{\Delta})$. For
ease of notation, we work in only one chiral sector of the worldsheet theory
and take $C$ and $D$ to be holomorphic fields. A conformal transformation
$z\mapsto f(z)$ on $\Sigma$ mixes the pair of fields as
\beq\new{\begin{array}{l}
C(z)~\mapsto~\left(\frac{df^{-1}(z)}{dz}\right)^\Delta\,C\left(f^{-1}(z)\right)
\\D(z)~\mapsto~\left(\frac{df^{-1}(z)}{dz}\right)^\Delta\left[D
\left(f^{-1}(z)\right)+\log
\left(\frac{df^{-1}(z)}{dz}\right)\,C\left(f^{-1}(z)\right)\right]\end{array}}
\label{CDmixing}\eeq
where an appropriate normalization for the $D$ operator has been chosen. It
follows that the corresponding states $|C\rangle=C(0)|0\rangle$ and
$|D\rangle=D(0)|0\rangle$ generate
a rank 2 Jordan cell structure for the Virasoro operator $L_0$,
\beq\new{\begin{array}{l}
L_0|C\rangle~=~\Delta|C\rangle\\L_0|D\rangle~=~\Delta|D\rangle+|C\rangle
\end{array}}
\label{Jordancell}\eeq
Consistency of invariance under the action of the Virasoro generators leads to
a set of partial differential equations \cite{r-tak} (or fusion rules
\cite{gurarie}) for the two- and three-point functions involving the
logarithmic pair $C,D$. For the two-point functions this yields
\cite{gurarie,r-tak}
\beq\new{\begin{array}{l}
\langle C(z)C(w)\rangle~=~0\\\langle C(z)D(w)\rangle~=~\frac a{(z-w)^{2
\Delta}}\\\langle D(z)D(w)\rangle~=~\frac1{(z-w)^{2\Delta}}\Bigl(b-2a
\log(z-w)\Bigr)\end{array}}
\label{2ptfns}\eeq
where $b$ is an arbitrary (integration) constant and the coefficient $a$ is
fixed by the leading logarithmic term in the conformal blocks.
These properties are the basic, non-trivial characteristics of quantum field
theories involving logarithmic operators, and are what will lead to some
noteworthy features in connection with their behaviour in moduli space.

Because of its logarithmic scaling violations, the quantum field theory
naturally contains an ultraviolet scale $\Lambda$ on $\Sigma$.\footnote{We
assume that all correlators of the quantum field theory are infrared soft and
ignore possible infrared divergences.} We suppose that the logarithmic pair
$C,D$ is constructed from the fields of some conformally-invariant theory with
action $S_*$. We may then consider the deformation of this theory by the
logarithmic operators defined by the action
\beq
S=S_*+\int_\Sigma d^2z~\Lambda^{2-\Delta-\overline{\Delta}}\left(g^CC(z,\bar
z)+g^DD(z,\bar z)\right)
\label{defaction}\eeq
where $g^C$ and $g^D$ are dimensionless coupling constants which depend on the
scale $\Lambda$. In writing \eqn{defaction} we have multiplied
together the holomorphic and antiholomorphic components of the logarithmic
fields. We are interested in the flows of these coupling constants under the
renormalization group, which in this case are non-trivial. In the case that
the deformation is given by operators which transform covariantly under
conformal transformations, the conformal field theory $S_*$ can be regarded as
an ultraviolet (or infrared when $\Sigma$ is compact) fixed point of this flow.
In the present case, however, the
logarithmic scaling violation \eqn{CDmixing} implies that the running coupling
constants transform non-covariantly under scale transformations
$z\mapsto\lambda z$,
\beq\new{\begin{array}{l}
g^C~\mapsto~g^C+(\log\lambda)\,g^D\\g^D~\mapsto~g^D\end{array}}
\label{gCDtransfs}\eeq

The coupling constants $(g^C,g^D)$ can be regarded as coordinates on the
moduli space of deformed conformal field theories of the form \eqn{defaction},
with origin taken at $S_*$. The Zamolodchikov metric on this space is defined
in a neighbourhood of $S_*$ by the two-point functions
\beq
G_{IJ}(S_*)\equiv(z-w)^{2\Delta}\,\bigl\langle I(z)J(w)\bigr
\rangle\Bigm|_{z-w=\Lambda}~~~~~~,~~~~~~I,J=C,D
\label{ZammetricCD}\eeq
which using \eqn{2ptfns} gives
\beq
G(S_*)=\pmatrix{0&a\cr a&b+2a t\cr}
\label{GmatrixCD}\eeq
where $t=-\log\Lambda$ is the renormalization group evolution parameter. The
full metric, constructed from both chiral components of the fields, factorizes
into holomorphic and antiholomorphic pieces each of the form \eqn{GmatrixCD}.
Now we change from the $C,D$ basis in coupling constant space to
\beq\new{\begin{array}{l}
g'^C~=~g^C-t\,
g^D\\g'^D~=~g^D\end{array}}
\label{gchange}\eeq
Then from \eqn{gCDtransfs} it follows that the new set of moduli space
coordinates \eqn{gchange} are (classically) invariant under scale
transformations.

To solve the corresponding renormalization group equations, we identify the
bare coupling constants $g_0'^I=g'^I(t=\infty)$ and expand the running coupling
constants $g'^I(t)$ as a power series in $g_0'^I$ near the critical point
$S_*$,
\beq
g'^I(t)=\e^{(2-\Delta)t}\,g_0'^I+\frac\pi{2-\Delta}\left(\e^{2(2-\Delta)t}-
\e^{(2-\Delta)t}\right)\sum_{J,K}c^I_{~JK}(S_*)\,g_0'^Jg_0'^K+{\cal O}(g_0'^3)
\label{g'tseries}\eeq
where $c_{~JK}^I$ are the operator product expansion coefficients of the
logarithmic pair at $S_*$ defined by
\beq
I(z)J(w)\sim(z-w)^{-\Delta}\,\sum_Kc^K_{~IJ}(\log\Lambda)
\,K\left(\mbox{$\frac12(z+w)$}\right)
\label{OPEIJ}\eeq
for $z-w\sim\Lambda$. The two-loop $\beta$-functions for the couplings
\eqn{gchange} are therefore given by the standard expression
\beq
\beta'^I[g';t]\equiv\frac{dg'^I(t)}{dt}=(2-\Delta)g'^I(t)-\pi\sum_{J,K}
c^I_{~JK}(S_*)\,g'^J(t)g'^K(t)
\label{beta2loop}\eeq
Note that the operator product expansion coefficients \eqn{OPEIJ} contain
logarithmic scaling violations, so that \eqn{beta2loop} also depends explicitly
on the evolution parameter $t$, in contrast to the usual Gell Mann-Low
$\beta$-functions in the scale-invariant case~\cite{mms}. They have been
computed in \cite{r-tak} and are given by
\beq\new{\begin{array}{lll}
c^C_{~CC}(t)&=&\mu+\nu
t\\c^D_{~CC}(t)&=&\nu\\c^C_{~CD}(t)&=&\tau+(\mu-\rho)t-\nu
t^2\\c^D_{~CD}(t)&=&\rho+t\\c^C_{~DD}(t)&=&\gamma+(2\tau-\delta)t+(\mu-2\rho)
t^2-\nu t^3\\c^D_{~DD}(t)&=&\delta+2\rho t+\nu t^2\end{array}}
\label{OPEcoeffs}\eeq
where $\mu,\nu,\dots$ are arbitrary (in general scale-dependent) constants.
Substituting \eqn{OPEcoeffs} into \eqn{beta2loop} and transforming back to the
unprimed coordinate system using \eqn{gchange} gives the two-loop
$\beta$-functions for the model \eqn{defaction}
\beq\new{\begin{array}{lll}
\beta^C[g;t]&=&(2-\Delta)\gc-\gd-\pi \mu (\gc)^2 - 2\pi\left[\tau + (1
-3\nu)t^2\right]\gc \gd\\& &~~~~~~~~~~ -\,\pi\left[\gamma+4\tau t - (1 -
4\nu)t^3\right](\gd)^2 \\
\beta^D[g;t]&=& (2 - \Delta ) \gd - \pi \nu (\gc)^2 -  2\pi\left[\rho +
(1-\nu)t\right]\gc\gd\\& &~~~~~~~~~~-\,\pi\left[\delta -
2(1-\nu)t^2\right](\gd)^2\end{array}}
\label{betafunctions}\eeq

As pointed out in \cite{mms}, the $\beta$-functions as they stand cannot by
themselves generate a gradient flow for some scalar function ${\cal C}$,
because in the moduli space $\beta^I$ are contravariant vectors whereas the
gradient flow $\partial{\cal C}/\partial g^I$ is a covariant vector. One needs
to contract the $\beta$-functions with the Zamolodchikov metric
\eqn{ZammetricCD} on moduli space to generate the appropriate gradient flow. To
put it differently, the operator product expansion coefficients $c^{I}_{~JK}$
are not completely symmetric in their indices, as would be required for the
existence of the function $\cal C$. In \cite{mms} it was shown that precisely
the contraction with the Zamolodchikov metric makes the covariant operator
product expansion coefficients $c_{IJK}=\sum_LG_{IL}c^L_{~JK}$ completely
symmetric in their indices. The gradient flow equation in a neighbourhood of
the fixed point $S_*$ is therefore written as
\be
\frac{\partial{\cal C}}{\partial g^I} = \sum_JG_{IJ}(S_*)\,\beta ^J
\label{gradient}\ee
To two-loop order, which yields the exact result within the Wilson
renormalization scheme, the Zamolodchikov metric contains no linear terms in
the $g^I$. This can be interpreted \cite{mms} as a choice of a normal
coordinate frame in moduli space, in which connection coefficients vanish.

Using \eqn{GmatrixCD} and \eqn{betafunctions} in \eqn{gradient}, we can now
write down the curl-free condition for the function $\cal C$ in the normal
coordinate frame. From this it follows that a gradient flow up to ${\cal
O}(g^2)$ can be satisfied provided that the operator product expansion
coefficients \eqn{OPEcoeffs} satisfy the conditions
\beq\new{\begin{array}{lll}
a(\rho - \mu) &=& b\nu  - a(1- 3\nu)t\\a(\delta - \tau ) &=& b\rho + [b(1-\nu)
+ 2a\rho]t + a(5-7\nu)t^2\end{array}}
\label{conditions}\eeq
Notice that these conditions necessitate the scale dependence of at least some
of the coefficients. For instance, it is not possible to arrange for a
simultaneous vanishing of the $t^2$ and $t$ dependent terms. With the
integrability conditions \eqn{conditions}, the flow equation \eqn{gradient} can
be integrated to yield
\beq\new{\begin{array}{lll}
{\cal C}[g;t]&=&c_*(t)+(2-\Delta)a\,\gc \gd -\mbox{$\frac{\pi}{3}$}\,a\nu
(\gc)^3 - a\pi\left[\rho + (1-\nu)t\right](\gc)^2\gd\\& &~~~~-\,
a\pi\left[\delta - 2(1-\nu)t^2\right]\gc (\gd)^2 + \mbox{$\frac{1}{2}$}
\left[(b+2at)(2-\Delta)-a\right](\gd)^2\\& &~~~~
-\,\mbox{$\frac{\pi}{3}$}\left[\gamma a + 4\tau a t - (1-4\nu)at^3
+(b+2at)(\delta - 2(1-\nu)t^2)\right](\gd)^3 \end{array}}
\label{cfunction}\eeq
where $c_*(t)$ is an arbitrary function of $t$, but not of the coupling
constants, arising from the moduli space integration. It can be set to a
constant by an appropriate choice of renormalization scheme. We note from
\eqn{conditions} that $a=0$ implies $b=0$, for which the effective action
\eqn{cfunction} is trivial. In general one needs the mixing between the pair of
logarithmic operators to generate a non-trivial gradient flow. In fact, the
coefficient $b$ can be set to 0 by a shift $D\to D+({\rm const.})\cdot C$ in
the logarithmic conformal algebra,  but we shall now see that invariance under
renormalization group transformations imposes much more severe restrictions on
the coefficients in \eqn{2ptfns}.

Notice that as a result of the explicit dependence on $t$, which cannot be
absorbed into a renormalization of the coupling constants, the flow function
$\cal C$ is in general not renormalization group invariant, in contrast to the
conventional cases. As we have mentioned, this is a generic feature of
logarithmic conformal field theory. Its total scale derivative is
\be
    \frac{d{\cal C}[g;t]}{dt}=\frac{\partial{\cal C}[g;t]}{\partial t} +
\sum_I\beta^I[g;t]\frac{\partial{\cal C}[g;t]}{\partial g^I}\ne 0
\label{flowrgni}
\ee
However, it is possible to cast ${\cal C}[g;t]$ in a form compatible with
standard renormalization group invariant quantities. This is precisely the case
in the conventional scale invariant two-dimensional field
theories~\cite{zam,mms}, where the function
$\cal C$ can be constructed out of renormalization group invariant two-point
functions \eqn{Cfundef} of the energy-momentum tensor\footnote{The fact that
the energy-momentum tensor is not renormalized stems from the renormalizability
of two-dimensional field theories.}. For the function (\ref{cfunction}), the
requirement $\partial{\cal C}/\partial t =0$ along with the integrability
conditions \eqn{conditions} are satisfied if and only if
\beq\new{\begin{array}{rrl}
\mbox{$\frac
d{dt}$}\,a&=&\mbox{$\frac{d}{dt}$}\,\nu~=~0\\\rho(t)&=&(\nu-1)t\\\delta(t)
&=&2(1-\nu)t^2\\b(t)&=&-2at\\\mu(t)&=&0\\\tau(t)&=&(3\nu-1)t^2\\\gamma(t)
&=&(5-16\nu)t^3
\end{array}}
\label{generalcond}\eeq
In \eqn{generalcond} we have ignored irrelevant integration constants
for simplicity. Their inclusion does not qualitatively affect the results and
can be easily incorporated below by adding \eqn{cfunction} for $t=0$. Then, the
renormalization group invariant $C$-function is
\beq
{\cal
C}_0[g]=c_*+(2-\Delta)a\,g^Cg^D-\mbox{$\frac\pi3$}\,a\nu(g^C)^3-\mbox{$\frac
a2$}\,(g^D)^2
\label{Ctindep}\eeq
For these choices the Zamolodchikov metric is off-diagonal, $G_{CC}=G_{DD}=0$,
$G_{CD}=G_{DC}=a$, and scale independent.

Note that this $t$ independence requirement fixes the operator product
expansion coefficients \eqn{OPEcoeffs} up to the constants $a$ and $\nu$. When
$\nu=0$, the only non-vanishing coefficient \eqn{OPEcoeffs} is
\beq
    c^C_{~DD} = t^3
\label{cdd}\ee
In the case that the deformation is marginal, i.e. $\Delta=2$, the
renormalization group equations can be integrated to yield
\be
    g^D(t) ={\rm const.}~~~~~~,~~~~~~g^C(t)={\rm const.}+g^D t -
\mbox{$\frac{\pi}{2}$}\, (g^D)^2 t^4
\label{gdgc}\ee
and the two-loop flow function is purely quadratic in the bare coupling
constant $g_0^D$,
\be
{\cal C}_0[g]^{\rm marg} = c_* -\mbox{$\frac{a}{2}$}\,(g^D_0)^2
\label{cfunction2}\ee

It is instructive at this stage to consider a specific example. We shall
briefly describe the situation for the case of D-brane recoil in string theory
\cite{recoil2}--\cite{lizzi}. Consider a set of bosonic fields $x^\mu(z,\bar
z)$ defined on a disc $\Sigma$, where $\mu=0,1,\dots,9$. These fields propagate
according to the free $\sigma$-model action $S_*=\frac12\int_\Sigma(\partial
x^\mu)^2$. The recoil of a D-particle in this model, due to the emission or
scattering of open and closed string states off of it, is described by the pair
of operators
\be
C_\epsilon=\epsilon\,\Theta _\epsilon (x^0)~~~~~~,~~~~~~ D_\epsilon=
x^0\,\Theta _\epsilon(x^0) \label{recoilop}\ee
where the infinitesimal parameter $\epsilon\to0^+$ regulates the step function
$\Theta(x^0)$. The operators \eqn{recoilop} ensure that, in an impulse
approximation, the D0-brane starts moving only at time $x^0=0$. They generate a
logarithmic conformal
algebra with conformal dimension $\Delta=-\epsilon^2/2<0$ \cite{recoil2}. The
total dimension of the corresponding deformation operator
$\int_\Sigma\partial\cdot[(y_iC_\epsilon+u_iD_\epsilon)\partial x^i]$,
$i=1,\dots,9$, added to the action $S_*$ is then $2-\epsilon^2$ which for
non-zero $\epsilon$ describes a relevant deformation of this conformal field
theory. It becomes marginal in the limit $\epsilon\to0^+$. Furthermore, in the
correlated limit $\epsilon\to0^+$ with
\beq
\epsilon^2\,t=-\epsilon^2\log\Lambda={\rm const.}
\label{corrlimit}\eeq
the two-point correlation functions of the pair \eqn{recoilop} are
\cite{recoil2}
\beq\new{\begin{array}{lll}
\langle C_\epsilon(z)C_\epsilon(0)\rangle&\sim&{\cal O}(\epsilon^2)\\\langle
C_\epsilon(z)D_\epsilon(0)
\rangle&\sim&\mbox{$\frac\pi2\sqrt{\frac\pi{\epsilon^2t}}$}
\,\Bigl(1-2\log(z/\Lambda)\Bigr)\\\langle
D_\epsilon(z)D_\epsilon(0)\rangle&=&\mbox{$\frac1{\epsilon^2}$}\,\langle
C_\epsilon(z)D_\epsilon(0)\rangle\end{array}}
\label{example}\eeq
Note that the last correlator in \eqn{example} follows from the general
(formal) property $D=\partial C/\partial\Delta$ of logarithmic conformal field
theories \cite{r-tak,W}. It is also straightforward to show that in the limit
$\epsilon\to0^+$ all the operator product expansion coefficients vanish except
that associated with the three-point function $\langle DDD\rangle$, which
diverges as $\epsilon\to0^+$.

Thus in the limit \eqn{corrlimit}, we recover a canonical logarithmic conformal
algebra, with the renormalization group invariant choice $b=-2at$ naturally
recovered from the elimination of the target space regularization parameter
$\epsilon$. In the present case, renormalization group invariance thus
translates into a relationship between target space and worldsheet scales. The
coupling constants $g^C$ of $C$ represent the collective coordinates of the
D-particle, while the couplings $g^D$ represent its momentum. The scale
parameter $t$ can be identified with the worldsheet zero-mode of the temporal
embedding field $x^0$. From (\ref{gdgc}) it follows that upon choosing an
appropriate renormalization scheme with new scale $t'=t-\frac{\pi}{2}t^4
g^D$, the coupling constant $g^D$ plays the r\^ole of a uniform velocity $u$
for the D-particle whose collective coordinate $g^C\sim y(t')$ follows the
linear time trajectory $y(t')=y+ut'$ corresponding to a Galilean boost of its
rest frame. The resulting quadratic action (\ref{cfunction2}) is the leading
order term of the standard Born-Infeld Lagrangian for D-particles. These
results have been established directly using $\sigma$-model perturbation theory
in \cite{lizzi}.

We now address the problem of representing the flow function obtained above in
terms of correlation functions of the quantum field theory, as in
\eqn{Cfundef}. The original ingredients of the proof of the $C$-theorem
\cite{zam} are the
conservation of the worldsheet energy-momentum tensor $T$, arising from
two-dimensional general coordinate invariance, and its non-renormalization. As
shown in \cite{mm}, the explicit forms of the two-point correlation functions
of the energy-momentum tensor are not important in this proof. The conservation
law for $T$ reads
\be
{\overline \partial}\,T_{zz}(z,\bar z) +\mbox{$\frac14$}\,\partial\,T_{z\bar
z}(z,\bar z) =0
\label{stresscons}\ee
Applying \eqn{stresscons} to the two-point functions in \eqn{Cfundef}, and
using the reality of these correlation functions which allows one to replace
the differential operators $z\partial$ and ${\bar z}{\overline \partial}$ by
$\partial/\partial t$, one easily arrives at the $C$-theorem~\cite{zam,mm}
\be
\frac{\partial\Phi[g]}{\partial t} = 12\int_\Sigma d^2z~z^2\bar
z^2\left\langle T_{z\bar z}(z,\bar z)T_{z\bar z}(0,0)\right\rangle_g
\label{ctheorem}\ee
without needing to explicitly know the components of the energy-momentum
tensor, but merely as a consequence of very general symmetries.

The important issue which arises in the presence of logarithmic deformations
concerns the renormalization of the energy-momentum tensor. We shall now argue
that the non-renormalization of $T$ extends to the logarithmic conformal field
theory case. For this, we consider the behaviour of the logarithmic pair $C,D$
under the action of the energy-momentum tensor field. From \eqn{CDmixing} it
follows that the pertinent operator product expansions read
\beq\new{\begin{array}{l}
T(z) C(w)~\sim~ \frac{\Delta}{(z-w)^2}\,C(w)+ \dots
\\T(z)D(w)~\sim~\frac{\Delta }{ (z-w)^2}\,D(w)+\frac{1}{(z-w)^2}\,C(w) + \dots
\end{array}}
\label{Tope}\eeq
for $z-w \rightarrow 0$, where the dots denote sub-leading terms. Using
\eqn{CDmixing} for a scale transformation $z\mapsto\e^{-t}z$ as $t$ varies, we
have the field scaling property
\beq
\frac{\partial C}{\partial t}=0~~~~~~,~~~~~~\frac{\partial D}{\partial t}=C
\label{CDscale}\eeq
Differentiating both sides of (\ref{Tope}) with respect to $t$ and using
\eqn{CDscale} it follows that
\be
  \frac\partial{\partial t}\left[(z-w)^2T(z)\right]C(w)= \frac\partial{\partial
t}\left[(z-w)^2T(z)\right]D(w)= 0
\label{nonscale}\ee
which implies that $T$ is not renormalized in the logarithmic conformal field
theory. Thus the standard proof of the $C$-theorem applies in this case,
allowing us to write (\ref{ctheorem}) in the form
\be
\sum_I\beta^I\frac{\partial\Phi}{\partial
g^I}=-12\,\sum_{I,J}\beta^IG_{IJ}\beta^J
\label{bexpre}\ee
where we have used the standard expression for the representation of
correlators of the energy-momentum tensor in terms of the deformation vertex
operators\footnote{This representation extends to the logarithmic case as a
consequence of the scale-invariance of the free energy of generic
two-dimensional quantum field theories.}.

{}From (\ref{gradient}) and \eqn{bexpre} it follows that the flow function
${\cal C}$ is related to the canonical Zamolodchikov $C$-function by
\be
\frac{\partial{\cal C}[g;t]}{\partial
g^I}=-\frac1{12}\,\frac{\partial\Phi[g]}{\partial g^I} +{\cal
A}_I[g;t]~~~~~~~~~~{\rm where}~~\sum_I{\cal A}_I[g;t] \beta^I[g;t] =0
\label{constr}\ee
The appearence of the function ${\cal A}_I$ in \eqn{constr} again implies
certain constraints among the operator product expansion coefficients of the
logarithmic conformal field theory, arising from the explicit scale dependences
of the quantities involved. For the examples of string $\sigma$-models in which
$\beta^I=\sum_JG^{IJ}\delta S/\delta g^J$, one may take it to be of the form
${\cal A}_I[g;t]=\sum_J\delta g^J{\cal K}_{JI}[g;t]$, where $\delta g^I$ is a
symmetry transformation of the coupling constants. The explicit form of the
function \eqn{Cfundef} has been computed in \cite{r-ts} and is given to leading
orders by
\beq
\Phi[g]=\phi_*-6\left[2a(2-\Delta)\,g^Cg^D-\Bigl(a-b_*(2-\Delta)\Bigr)
\,(g^D)^2\right]
\label{Phiexpl}\eeq
which yields the scale-invariant flow function \eqn{Ctindep} to ${\cal O}(g^2)$
in the coupling constant expansion\footnote{The constant coefficient $b_*$ is
the scale invariant part of the condition $b(t)=-2at+b_*$ in
(\ref{generalcond}), which was ignored in the analysis above for simplicity.
{}From (\ref{cfunction})
and (\ref{Phiexpl}) it is straightfoward to see that the inclusion of such
integration constants in the conditions (\ref{generalcond}) does not affect the
identification of $-\frac{1}{12}\Phi$ with the renormalization group invariant
flow function ${\cal C}_0[g]$ to ${\cal O}[g^2]$.}. Note that this need not
imply that the usual $C$-theorem holds because the right-hand side of
\eqn{ctheorem} need not be negative for non-unitary models, as is the case of
most logarithmic conformal field theories. Some conditions for the monotonicity
of \eqn{Phiexpl} under a change of scale are given in \cite{r-ts}.

It is straightforward to generalize the above analysis to the case where the
logarithmic conformal field theory contains a rank $n>2$ Jordan block, as well
as more than one Jordan cell structure, using the two-point correlation
functions and operator product expansion coefficients computed in \cite{r-tak}.
One problem not addressed by the present analysis is the existence of the
gradient flow {\it globally} on the moduli space, away from a neighbourhood of
a critical point $S_*$. There may be topological obstructions in moduli space
to the flow, and it would be interesting to examine more precisely the global
properties of this space for logarithmic conformal field theories. We note that
the explicit scale-dependence in the operator product expansion
coefficients of a logarithmic conformal field theory implies the breakdown of
their interpretation as factorizable S-matrix elements in target space. An
analogous situation is present in the case of open systems, where an explicit
renormalization group dependence arises from ultraviolet divergences at the
turning points of closed time-like paths, and also in Liouville (non-critical)
string theory where the correlators depend explicitly on the worldsheet area in
the fixed-area formalism~\cite{area}. In this latter case the Liouville field
can be identified with the target space time coordinate \cite{emn}, and the
Liouville deformations of the cosmological constant are given by a set of
logarithmic operators.


\begin{thebibliography}{99}

\baselineskip=12pt

\bibitem{gurarie} L. Rozansky and H. Saleur, Nucl. Phys. {\bf B376} (1992)
461;\\ V. Gurarie, Nucl. Phys. {\bf B410} [FS] (1993) 535;\\ M.A.I. Flohr, Int.
J. Mod. Phys. {\bf A11} (1996) 4147; {\bf A12} (1997) 1943;\\ M.R. Gaberdiel
and H.G. Kausch, Nucl. Phys. {\bf B489} (1996) 293; Phys. Lett. {\bf B386}
(1996) 131;\\ F. Rohsiepe, {\it On Reducible but Indecomposable Representations
of the Virasoro Algebra}, hep-th/9611160;\\ I.I. Kogan, A. Lewis and O.A.
Soloviev, Int. J. Mod. Phys. {\bf A13} (1998) 1345.

\bibitem{cm} X.-G. Wen, Y.-S. Wu and Y. Hatsugai, Nucl. Phys. {\bf B422} [FS]
(1994) 476;\\ J.S. Caux, I.I. Kogan and A.M. Tsvelik, Nucl. Phys.
{\bf B466} (1996) 444;\\ G.M. Watts, {\it A Crossing Probability for Critical
Percolation in Two-dimensions}, cond-mat/9603167;\\ M.A.I. Flohr, Mod. Phys.
Lett. {\bf A11} (1996) 55; \\ Z. Maassarini and D. Serban, Nucl. Phys. {\bf
B489} (1997) 603;\\ V. Gurarie, M.A.I. Flohr and C. Nayak, Nucl. Phys. {\bf
B498} (1997) 513;\\ J.S. Caux, N. Taniguchi and A.M. Tsvelik, Phys. Rev. Lett.
{\bf 80} (1998) 1276.

\bibitem{recoil} A. Bilal and I.I. Kogan, Nucl. Phys. {\bf B449} (1995) 569;\\
I.I. Kogan and N.E. Mavromatos, Phys. Lett. {\bf B375} (1996) 111;\\ V. Periwal
and O. Tafjord, Phys. Rev. {\bf D54} (1996) 4690;\\ D. Berenstein, R. Corrado,
W. Fischler, S. Paban and M. Rozali, Phys. Lett. {\bf B384} (1996) 93;\\ I.I.
Kogan and A. Lewis, Nucl. Phys. {\bf B509} (1998) 687.

\bibitem{recoil2} I.I. Kogan, N.E. Mavromatos and J.F. Wheater, Phys. Lett.
{\bf B387} (1996) 483.

\bibitem{st} J. Ellis, N.E. Mavromatos and D.V. Nanopoulos, Int. J. Mod. Phys.
{\bf A12} (1997) 2639.

\bibitem{lizzi} F. Lizzi and N.E. Mavromatos, Phys. Rev. {\bf D55} (1997) 7869.

\bibitem{nl} M.R. Rahimi Tabar and S. Rouhani, Ann. Phys. {\bf 246} (1996) 446;
Nuovo Cimento {\bf B112} (1997) 1079; Phys. Lett. {\bf A224} (1997) 331;
Europhys. Lett. {\bf 37} (1997) 447;\\ M.A.I. Flohr, Nucl. Phys. {\bf B482}
(1996) 567;\\ O. Coceal, W.A. Sabra and S. Thomas, {\it Conformal Solutions of
Duality Invariant $2D$ Magnetohydrodynamic Turbulence}, hep-th/9604157.

\bibitem{zam} A.B. Zamolodchikov, JETP Lett. {\bf 43} (1986) 731; Sov. J. Nucl.
Phys. {\bf 46} (1987) 1090; A.A.W. Ludwig and J.L. Cardy, Nucl. Phys. {\bf
B285} [FS19] (1987) 687.

\bibitem{tseytl} A.A. Tseytlin, Phys. Lett. {\bf B194} (1987) 63;\\ H. Osborn,
Phys. Lett. {\bf B214} (1988) 555.

\bibitem{mm} N.E. Mavromatos and J.L. Miramontes, Phys. Lett. {\bf B212} (1988)
33;\\ N.E. Mavromatos, Phys. Rev. {\bf D39} (1989) 1659.

\bibitem{r-ts} M.R. Rahimi-Tabar and S. Rouhani, {\it Zamolodchikov's C-theorem
and the Logarithmic Conformal Field Theory}, hep-th/9707060.

\bibitem{mms} N.E. Mavromatos, J.L. Miramontes and J.M. S\'anchez de Santos,
Phys. Rev. {\bf D40} (1989) 535.

\bibitem{r-tak} M.R. Rahimi-Tabar, A. Aghamohammadi and M. Khorrami, Nucl.
Phys. {\bf B497} (1997) 555;\\ M.A.I. Flohr, Nucl. Phys. {\bf B514} (1998) 523.

\bibitem{W} A. Shafiekhani and M.R. Rahimi-Tabar, Int. J. Mod. Phys. {\bf A12}
(1997) 3723.

\bibitem{area} J. Ellis, N.E. Mavromatos and D.V. Nanopoulos, Mod. Phys. Lett.
{\bf A12} (1997) 1759.

\bibitem{emn} J. Ellis, N.E. Mavromatos and D.V. Nanopoulos, Phys. Lett. {\bf
B293} (1992) 37;\\ I.I. Kogan, in: {\it Particles and Fields '91}, eds. D.
Axen, D. Bryman and M. Comyn (World Scientific, Singapore, 1992) 837.

\end{thebibliography}
\end{document}